\documentstyle[12pt,aaspp4]{article}
\def\del{{\partial}}

\def\eps{\epsilon}

\def\sD{{\mathcal{D}}}
\def\sE{{\mathcal{E}}}
\def\spose#1{\hbox to 0pt{#1\hss}}
\def\lta{\mathrel{\spose{\lower 3pt\hbox{$\mathchar''218$}}
     \raise 2.0pt\hbox{$\mathchar''13C$}}}
\def\gta{\mathrel{\spose{\lower 3pt\hbox{$\mathchar''218$}}
     \raise 2.0pt\hbox{$\mathchar''13E$}}}
\lefthead{Gammie}
\righthead{MHD Accretion}
\begin{document}
%
%\submitted{Draft version \today}
%
\title{Efficiency of Magnetized Thin Accretion Disks in the Kerr Metric}
\author{Charles F. Gammie \altaffilmark{1}}
\affil{Department of Astronomy and Department of Physics\\
University of Illinois at Urbana-Champaign \\
1002 W. Green St., Urbana IL 61801; gammie@uiuc.edu}

\altaffiltext{1}{also Harvard-Smithsonian Center for Astrophysics,
MS-42, 60 Garden St., Cambridge, MA 02138, USA}

\today

\begin{abstract}

The efficiency of thin disk accretion onto black holes depends on the
inner boundary condition, specifically the torque applied to the disk at
the last stable orbit.  This is usually assumed to vanish.  I estimate
the torque on a magnetized disk using a steady magnetohydrodynamic
inflow model originally developed by \cite{tak90}.  I find that the
efficiency $\eps$ can depart significantly from the classical thin disk
value.  In some cases $\eps > 1$, i.e. energy is extracted from the
black hole.

\end{abstract}


\section{Introduction}

The dynamics of accretion close the event horizon of black holes is of
considerable interest to astronomers because that is where most of the
accretion energy is released and because strong-field gravitational
effects may be evident there.

Much of the work on thin disks around black holes assumes that the disk
ends at or near $r = r_{mso}$, the radius of the marginally stable
circular orbit.  Inside $r_{mso}$, material is assumed to nearly
free-fall onto the black hole (e.g.  \cite{bar70,tho74,ajs78}).  This
implies the so-called ``no-torque'' boundary condition on the disk, and
leads to a disk whose surface brightness vanishes at the inner edge
(\cite{ss73}).  The no-torque boundary condition leads to the classical
estimate for the accretion efficiency $\eps_0 = 1 - \sE(r_{mso})/c^2$,
where $\sE(r_{mso}) = -u_t(r_{mso})$ is the ``energy at infinity'' per
unit mass of a particle at the marginally stable orbit.  $\eps_{0}$
ranges between $1 - 2 \sqrt{2}/3 \approx 0.057$ for a nonrotating ($a =
0$) black hole, to $1 - 1/\sqrt{3} \approx 0.42$ for a prograde disk
around a maximally rotating ($a = 1$) black hole.  

It was realized early on, however, that magnetic fields might alter the
dynamics of the accreting material and hence the accretion efficiency
(see, e.g.,  J. M. Bardeen quoted in \cite{tho74}).  More recently, it
has been argued that the magnetic fields threading the accreting
material should be strong enough to be dynamically important, but not so
strong that the fields can be regarded as force-free (\cite{kro99}).

In this Letter I estimate the magnetic torque on the inner edge of a
thin accretion disk in the Kerr metric.  I use a model for the inflow
that is based on earlier work by \cite{tak90}, following \cite{phi83}
and Camenzind 1986a,1986b,1989.

\section{Model}

Consider a thin disk in the equatorial plane of the Kerr metric.
Because the disk is thin, $c_s^2/c^2 \ll 1$, so the relativistic
enthalpy $\eta \approx 1$.  The disk is magnetically turbulent
(\cite{bh91}) with magnetic energy density $B^2/(8\pi) \approx 2\alpha
\rho c_s^2$ (\cite{hgb95}), where $\alpha$ is the usual dimensionless
viscosity of accretion disk theory.  I will assume that $\alpha \ll 1$
so that magnetic fields make a negligible contribution to the
hydrostatic equilibrium of the disk.   Then the disk has an inner
surface at $r = r_{in}$, with $r_{mso} - r_{in} \sim r_{mso} (c_s/c)^2$,
so $r_{in} \approx r_{mso}$.  At the inner surface of the disk I imagine
that magnetic field lines rise approximately radially through the disk
atmosphere and force the atmosphere to corotate with the disk surface.
High enough in this corotating atmosphere, the gradient of the effective
potential changes sign and the gas begins to stream inward along field
lines toward the black hole.  As gas leaves the disk, the character of
the flow changes, becoming less turbulent and more nearly laminar.

This picture leads one to consider a steady, axisymmetric inflow close
to the equatorial plane of the Kerr metric.  In what follows, I will
work exclusively in Boyer-Lindquist coordinate $t,r,\theta,\phi$, and
follow the notational conventions of \cite{mtw}.  I assume that, in the
poloidal plane, the fluid velocity and magnetic field are purely radial,
i.e.  that $u^\theta \equiv 0$ and $F_{r\phi} \equiv 0$, where $u^{\mu}$
is the four-velocity and $F_{\mu\nu}$ is the electromagnetic field
(``Maxwell'') tensor; recall that in the nonrelativistic limit, denoted
$\rightarrow$, $F_{r\phi}(\theta = \pi/2) \rightarrow r B_\theta$, where
${\bf B}$ is the magnetic field three-vector.  I also assume that all
flow variables are functions only of $r$, i.e., that $\del_\theta = 0$.
This one-dimensional flow is similar in spirit to the old \cite{wd67}
model for the solar wind, turned inside-out.  As in the Weber-Davis
model, the magnetic field has a monopolar geometry.

I will also assume perfect conductivity, so that the electric field vanishes
in a frame comoving with the fluid:
\begin{equation}\label{PERFCON}
 u^{\mu} F_{\mu\nu} = 0.
\end{equation}
Together with the symmetry conditions, this leaves $F_{\mu\nu}$ with six
nonzero components: $F_{t\theta}$ ($\rightarrow -r E_\theta$),
$F_{r\theta}$ ($\rightarrow -r B_\phi$), and $F_{\theta\phi}$
($\rightarrow -r^2 B_r$), and gives the relation $F_{r\theta} = -
(F_{t\theta} u^t - F_{\theta\phi} u^\phi)/u^r$.  Thus the
electromagnetic field has only two degrees of freedom.  

The flow symmetries reduce Maxwell's equations to
\begin{equation}
\del_r F_{t\theta} = 0
\end{equation}
and
\begin{equation}
\del_r F_{\theta\phi} = 0.
\end{equation}
The first equation is the relativistic ``isorotation'' law, which can be
rewritten $F_{t\theta} = \Omega_F F_{\theta\phi}$, where $\Omega_F$ is
the rotation frequency $u^\phi/u^t$ at the radius where $u^r = 0$.  The
second equation is the relativistic equivalent of $\nabla\cdot B = 0$.

Conservation of particle number leads to a conserved ``rest mass flux'',
per unit $\theta$,
\begin{equation}
F_M = 2\pi r^2 \rho u^r \rightarrow 2\pi r^2 \rho v_r,
\end{equation}
here $\rho \equiv$ particle number density multiplied by the rest mass
per particle and ${\bf v}$ is the three-velocity.  The conserved angular
momentum flux is
\begin{eqnarray}
F_L & = &  2\pi r^2 T_\phi^r = 2 \pi r^2 \left(\rho u_\phi u^r - {\sD
	F_{\theta\phi}
	F_{r\theta}\over{4\pi r^2}}\right) \\
	& & \rightarrow
	2\pi r^3 \big[\rho v_r v_\phi  - {B_\phi B_r\over{4\pi}}\big],
	\nonumber
\end{eqnarray}
where $T_{\mu\nu}$ is the stress-energy tensor, $\sD \equiv 1 - 2 r_g/r + a^2
r_g^2/r^2$, $r_g \equiv G M/c^2$.  The conserved mass-energy flux is
\begin{eqnarray}
F_E & = &  -2\pi r^2 T_t^r = 2 \pi r^2 \left(-\rho u_t u^r - {\sD
	F_{t\theta} F_{r\theta}\over{4\pi r^2}}\right) \\
	& & \rightarrow
	2\pi r^2 \Big[\rho v_r \big(c^2 + {1\over{2}}(v_r^2 + v_\phi^2) 
	- {G M\over{r}}\big)  \nonumber \\ 
	& & + v_r {B_\phi^2\over{4\pi}} - v_\phi {B_r B_\phi
	\over{4\pi}}\Big],
	\nonumber
\end{eqnarray}
and, consistent with the thin disk approximation, I have neglected the
thermal energy's contribution to $F_E$.  Notice that the accretion
efficiency is given by $\eps = 1 - F_E/F_M$.

The final equation needed to close the system is the normalization of
four-velocity, 
\begin{equation}\label{FOURNORM}
u^\mu u_\mu = -c^2.
\end{equation}
Henceforth I will set $r_g = c = -F_M = 1$. \footnote{Physical units may
be recovered as follows.  Length: $r_g$; time: $r_g/c$; mass: $-F_M
c/r_g$; $F_{\theta\phi}$: $G M (-F_M)^{1/2} c^{-3/2}$, while $F_M
\approx -\dot{M} r/(2 H)$.}

Equations (\ref{PERFCON}) through (\ref{FOURNORM}) describe a one
dimensional inflow model with a series of algebraic relations.  This
model is physically identical to that developed by \cite{tak90},
although cast is somewhat different notation.  Also, unlike
\cite{tak90}, I specialize to the case where the inflow is anchored in a
thin disk at the marginally stable orbit.  This sets the inflow in a
specific astrophysical context and allows one to relate it to an
accretion efficiency.  These inflow solutions have also been explored by
\cite{cam94,cam96} and, in unpublished work, by Camenzind and Englmaier.

\section{Boundary Conditions}

The inflow model has 6 dynamical variables: $\rho, u^t, u^r, u^\phi,
F_{t\theta},$ and $F_{\theta\phi}$, and 6 conserved quantities: $F_M,
F_L, F_E, F_{\theta\phi}, \Omega_F,$ and $u^\mu u_\mu$, so the equations
are fully integrated.  What sets the conserved quantities?

The mass flux $F_M$ has been normalized to $-1$, but more physically it
is set by conditions in the disk at large radius.  The normalization of
four-velocity gives $u^\mu u_\mu = -1$.  $F_{\theta\phi}$ is related to
the magnetic flux emerging from the inner edge of the disk.  This is
presumably determined by the action of magnetohydrodynamic (MHD)
turbulence in the disk and the interaction of the disk with the inflow
itself.  I will treat it as a parameter of the problem, and estimate
reasonable values for it later.  Three quantities remain to be
specified.

It seems reasonable to require that the energy and angular momentum
fluxes be continuous across the boundary between the inflow and the
disk.  It also seem reasonable to require that the inflow four-velocity
match continuously onto the disk.  Happily, these requirements do not
overconstrain the problem.  If $\Omega_F = \Omega(r_{in})$ and
$u^r(r_{in}) = 0$, then one can show that $F_E = \Omega F_L + (\sE - l
\Omega) F_M$ at $r_{in}$; here $l \equiv u_\phi$.  This eliminates two
degrees of freedom by expressing $F_E$ in terms of $F_L$ and by fixing
$\Omega_F$.  The energy and angular momentum flux are then continuous if
the viscous (more properly, turbulent) angular momentum flux of the disk
just outside $r_{in}$ matches the electromagnetic angular momentum flux
of the inflow just inside $r_{in}$: 
\begin{equation}
T_\phi^r({\rm visc}) = T_\phi^r({\rm EM}).  
\end{equation} 
This condition will be automatically satisfied once the disk relaxes
to a steady state, since the disk will increase its surface density
until it can carry off the entire outward electromagnetic angular
momentum (and energy) flux from the inflow.  

The final degree of freedom ($F_L$, or equivalently, $F_E$) is fixed by
the condition that the flow pass smoothly through the fast critical
point.  \footnote{The Alfv\'en point does not impose any new condition
on the flow, since all trans-Alfv\'enic solutions pass smoothly through
the Alfv\'en point; see, e.g., \cite{phi83}.  The slow point is absent
because the flow is cold.}  Thus $F_L$ and $F_E$ emerge as
``eigenvalues'' of the solution.

\section{Solutions}

Once the model parameters are fixed, the resulting set of nonlinear
algebraic equations  must be solved numerically.  I obtain $F_L$, and
the location of the fast critical point ($r_f, u^r_f$) via simultaneous
solution of 
\begin{eqnarray}
\del_{u^r} F_E(r_f,u^r_f,F_L) & = & 0, \\
\del_{r} F_E(r_f,u^r_f,F_L) & = & 0, \\
F_E(r_f,u^r_f,F_L) - F_E(r_{mso},0,F_L) & = & 0.
\end{eqnarray}
Here I have used the fact that the critical point is a saddle point of
$F_E(r,u^r; F_L)$.  I use the multidimensional Newton-Raphson method of
\cite{pre92}, and for simplicity I evaluate the derivatives numerically.

There is one subtlety involved in the solution.  In calculating
$F_E(r,u^r,F_L)$ one must solve a quadratic equation for $u^t$ (or
equivalently $u^\phi$), so one must decide which root is physical.  In
general this is a nontrivial matter, since the physical solution skirts
a region in the $r,u^r$ plane where the discriminant of the quadratic
vanishes (the solution makes a smooth transition from one branch of the
solution to another).  Fortunately it turns out that one branch is
appropriate in the neighborhood of $r_{mso}$ and the other in the
neighborhood of the fast point, so in practice this subtlety is easily
dealt with.

Figure 1 shows $u^r(r)$ for a solution with $a = 0.95$ and
$F_{\theta\phi} = 6$.  The dot marks the fast point at $(r,u^r) = (1.37,
-0.26)$.  This solution has $F_E \approx 0.04$ and $F_L \approx 1.23$,
so $\eps \equiv 1 + F_E \approx 1.04$, i.e. energy is extracted from the
black hole.  Evidently magnetic fields can significantly alter the
efficiency from the classical value.

Figure 2 shows contours of the eigenvalue $F_L$ in a survey over the
$a-F_{\theta\phi}$ plane (see also Fig. 3 of \cite{cam94}).  For strong
field and large black hole spin (up and to the right of the heavy solid
line in the figure), $F_L > 0$.  \cite{tak90} have shown that $F_L > 0$
if the field rotation frequency $\Omega_F$ is exceeded by the
characteristic rotation frequency of the space time, $2 a/(r^3 (1 +
a^2/r^2 + 2 a^2/r^3$)), at the Alfv\'en point.

Figure 3 shows contours of the accretion efficiency $\eps$ evaluated
over a portion of the $a-F_{\theta\phi}$ plane (see also Fig. 4 of
\cite{cam94}).  The contours in Figure 3 are located at intervals of
$\Delta \eps = 0.05$.  For strong field and large black hole spin (up
and to the right of the heavy solid line in the figure), $F_E > 0$, that
is, energy is extracted from the black hole.  As \cite{tak90} have
shown, this can only happen if the Alfv\'en point lies within the
ergosphere ($r < 2$).  The heavy dashed line corresponds to the
classical efficiency of prograde disk around a maximally rotating black
hole, $\eps \approx 0.42$.

\section{Astrophysical Discussion}

What is an astrophysically sensible value for the crucial magnetic field
strength parameter $F_{\theta\phi}$?  I will make a purely Newtonian
estimate, for clarity and because the relativistic corrections are
likely to be smaller than the other sources of uncertainty.  Suppose
that the radial field leaving the disk is $= f B_d$, where subscript $d$
denotes a quantity evaluated in the disk and $f \lesssim 1$.  It is a
result of numerical models of disk turbulence that $B_d^2/(8\pi) \approx
2\alpha\rho_d c_{s,d}^2$ (e.g.  \cite{hgb95}).  Using the usual steady state
disk equation $3\pi\Sigma\nu \approx \dot{M} \approx -2 F_M (H/r)$, I
find
\begin{equation}
F_{\theta\phi} \approx r^2 B_r \approx 2.3 f r_d^{3/4}.
\end{equation}
If the inner edge of the disk is at $r_d = 6$, $F_{\theta\phi} \approx
8.8 f$,  so the region of parameter space shown in Figures 2 and 3 is
likely relevant to disks.

The inflow model thus suggests that the presence of a modest magnetic
field, and the accompanying torque on the inner edge of the disk, can
significantly increase the efficiency of thin disk accretion onto black
holes.  For $a = 0$ and $F_{\theta\phi} \ll 1$, $\eps \approx \eps_{0} +
0.01 |F_{\theta\phi}|$, while for $F_{\theta\phi} = 4$, $\eps = 0.165$.
The added luminosity manifests itself as an increase in the surface
brightness of the disk due to the torque applied at its inner boundary.
In the same quasi-Newtonian spirit as the estimate of $F_{\theta\phi}$,
the surface brightness profile becomes $\propto 1 - \beta
(r_{mso}/r)^{1/2}$ (see \cite{st83}) and $\beta \approx 1$ in the
Shakura-Sunyaev model.  Here $\beta \equiv 1 - F_L({\rm EM})/l_{mso}
\lesssim 1$.

The inflow model also suggests that, because material accretes with a
smaller specific angular momentum than it would in the absence of
magnetic fields, it is more difficult to manufacture a rapidly spinning
hole by disk accretion.  An equilibrium spin is reached when $F_L = 2 a
F_E$ (e.g.  \cite{pg98}, eq.  10).  For $F_{\theta\phi} = 4$ this
equilibrium value is reached at $a \approx 0.7$.

A major limitation of this study (which should be regarded as an
instructive example rather than a source for estimating efficiencies),
is that I have ignored the vertical structure of the inflow.  Crudely
speaking, one might expect that field lines which do not lie in the
midplane are more lightly loaded ($|F_{\theta\phi}|$ is larger) so that
specific energy and angular momentum fluxes might increase away from the
midplane, until at sufficient latitude one reaches a field line which
would rather inflate away than remain tied between the inflow and the
disk.  At high latitude, then, the outward electromagnetic energy flux
might emerge in the form of a wind, and be better described by the
force-free magnetosphere model of \cite{bz77}.

Another major limitation is the simplified, monopolar field geometry.
This limitation can be overcome by direct numerical integration of the
basic equations.  Evidently numerical models of inflow inside the
marginally stable orbit would be enormously interesting.  Fortunately
they are now practical, at least within the MHD approximation.

\acknowledgments

Stu Shapiro's thoughtful advice has greatly improved this paper.  I am
also grateful to Eric Agol, Roger Blandford, Jeremy Goodman, John
Hawley, Julian Krolik, Jochen Peitz, and Gordon Ogilvie for their
comments, and to Phil Myers and the Radio Group at the Center for
Astrophysics for their hospitality.

\clearpage

\begin{figure}
\plotone{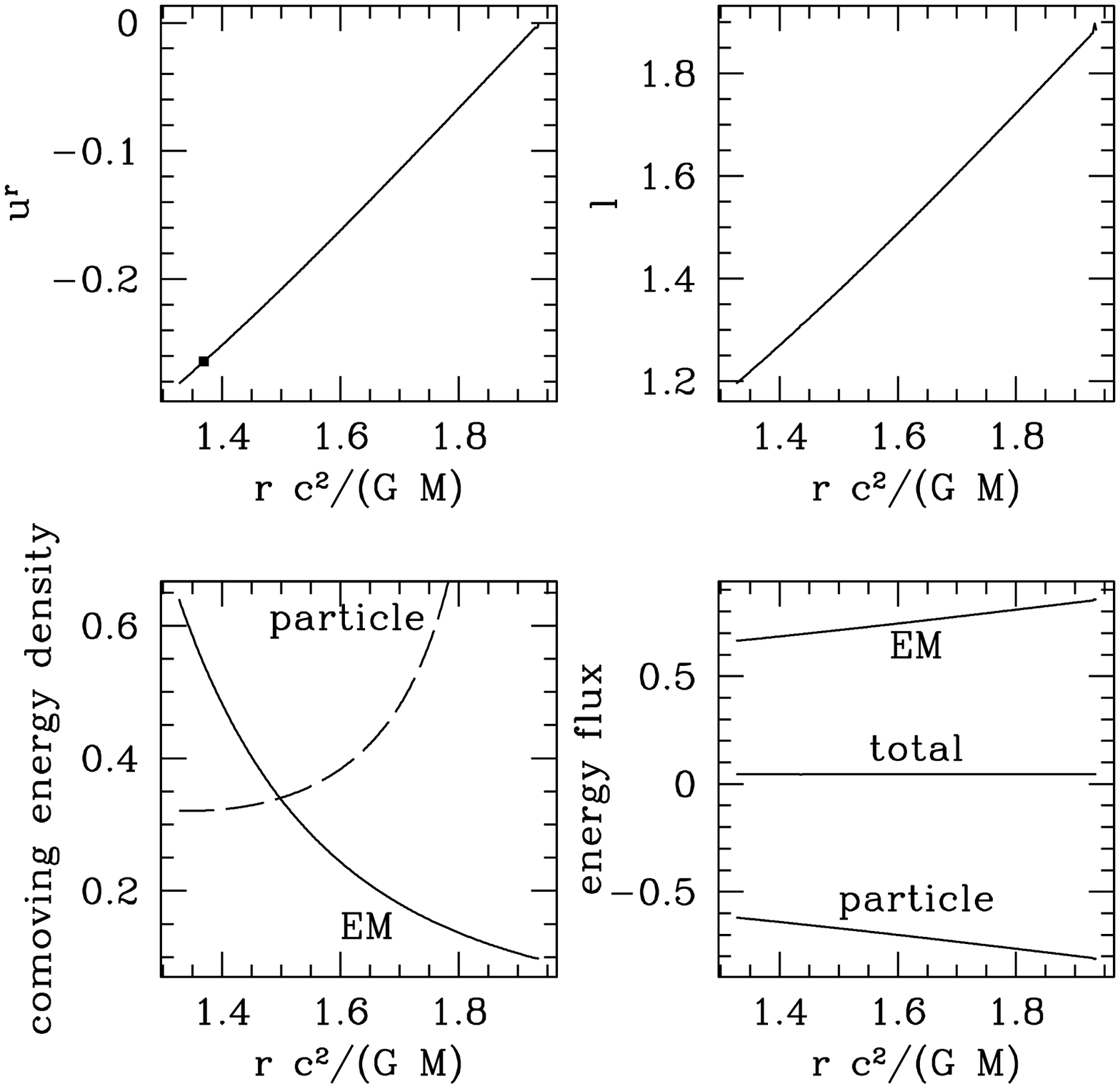}
\caption{
An example solution with $a = 0.95$ and $F_{\theta\phi} = 6$.  The upper
left panel shows $u^r(r)$, the upper right $l = u_\phi(r)$.  The dot
shows the location of the fast critical point.  The lower left panel
shows the electromagnetic energy density ($T_{\mu\nu}({\rm EM}) u^\mu
u^\nu$; solid line) and the rest-mass energy density of the accreting
particles ($\rho$; dashed line) measured in a frame comoving with the
fluid.  The lower right panel shows the electromagnetic, particle, and
total energy fluxes as a function of radius.
}
\end{figure}

\begin{figure}
\plotone{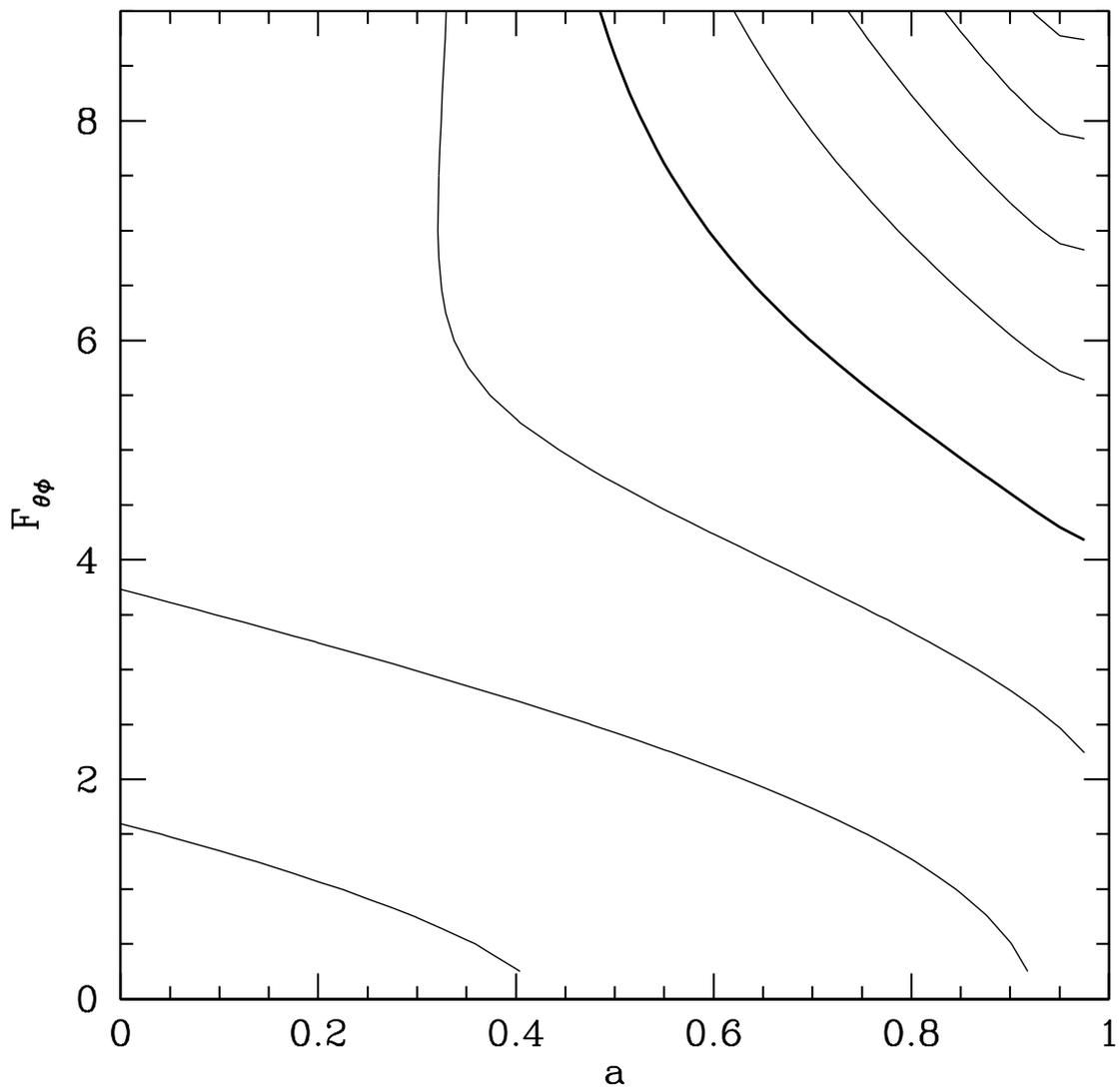}
\caption{
Contours of constant angular momentum flux $F_L$ in the $a,F_{\theta\phi}$
plane.  The zero angular momentum flux contour is marked as a heavy solid
line; above and to the right of this line $F_L > 0$.  The light contours are
linearly spaced at $\Delta F_L = 1$.
}
\end{figure}

\begin{figure}
\plotone{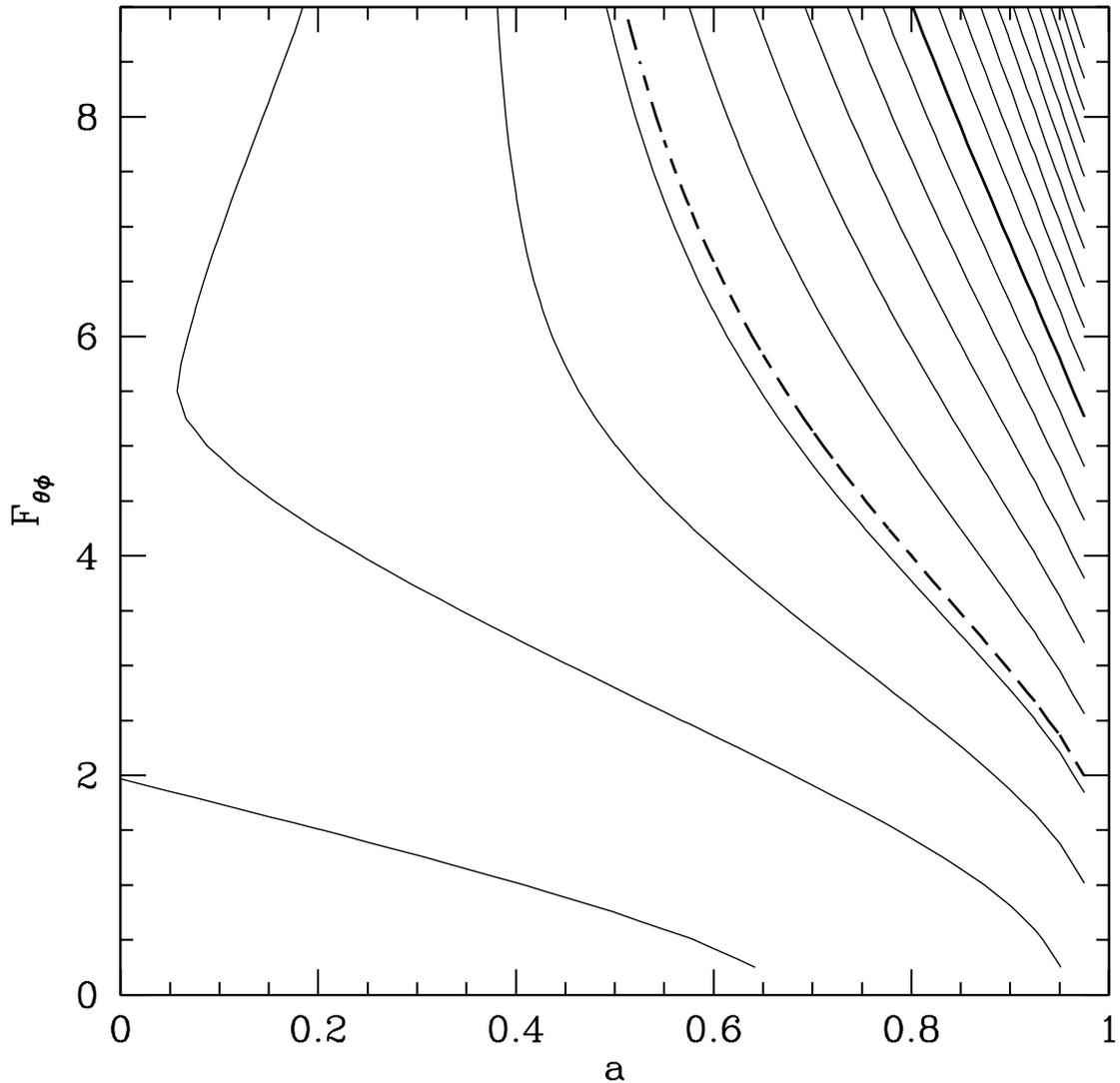}
\caption{
Contours of constant efficiency $\eps \equiv 1 + F_E$ in the
$a,F_{\theta\phi}$ plane.  The unit efficiency contour is marked as a
heavy solid line; above and to the right of this line $F_E > 0$, i.e.
energy is being extracted from the black hole.  The light contours are
linearly spaced at $\Delta \eps = 0.1$.  The heavy dashed contours lies
at $\eps \approx 0.42$, the classical efficiency of prograde disk around
a maximally rotating black hole.
}
\end{figure}

\end{document}